\documentclass[sigconf,natbib=true]{acmart}

\usepackage{booktabs}
\usepackage{multirow}
\usepackage{array}
\usepackage{tabularx}
\usepackage[table]{xcolor}
\usepackage{graphicx}
\usepackage{balance}
\usepackage{microtype}

\graphicspath{{figures/}}

\copyrightyear{2026}
\acmYear{2026}
\setcopyright{cc}
\setcctype{by}
\acmConference[CIKM '26] {Proceedings of the 35th ACM International Conference on Information and Knowledge Management}{November 7--11, 2026}{Rome, Italy.}
\acmBooktitle{Proceedings of the 35th ACM International Conference on Information and Knowledge Management (CIKM '26), November 7--11, 2026, Rome, Italy}
\acmISBN{979-8-4007-2539-5/2026/11}
\acmDOI{10.1145/3799682.3839970}

\begin{document}

\title{Beyond Benchmark Scores: How Synthetic and Authentic Query Distributions Diverge in RAG Evaluation}

\author{Filip J. Kucia}
\orcid{0009-0005-8473-1402}
\email{filip.kucia.dokt@pw.edu.pl}
\affiliation{%
  \department{Faculty of Mathematics and Information Science}
  \institution{Warsaw University of Technology}
  \city{Warsaw}
  \country{Poland}
}
\affiliation{%
  \institution{Sparrows AI sp.\ z o.o.}
  \city{Pu\l{}awy}
  \country{Poland}
}
\author{Barbara M. Gawlik}
\orcid{0009-0005-6029-8456}
\email{barbara.gawlik.stud@pw.edu.pl}
\affiliation{%
  \department{Faculty of Mathematics and Information Science}
  \institution{Warsaw University of Technology}
  \city{Warsaw}
  \country{Poland}
}
\renewcommand{\shortauthors}{Filip J. Kucia and Barbara M. Gawlik}
\begin{abstract}
RAG systems are routinely evaluated using synthetic question sets generated from the target document corpus. While this practice provides a useful check on overall retrieval capability, relying exclusively on synthetic benchmarks can mislead under distribution shift and overstate deployment readiness. Synthetic generation spreads questions evenly across the corpus, formulating long, detailed queries; real users put most of their traffic on a few administrative and procedural topics in short queries, while also asking about matters the generator never covers at all.
We demonstrate this gap on a university faculty information system, comparing 1,851 synthetic questions generated via Gemini Notebook against 322 authentic queries collected via a student survey. The synthetic and authentic query sets differ significantly: authentic queries average 6.8 words versus 15.7 for the synthetic ones, and draw from only 53 unique sources compared to 165. Consequently, configurations that appear highly effective on synthetic benchmarks experience a substantial performance drop on authentic queries. Importantly, optimizing on synthetic queries selected a higher-latency hybrid retriever. In our setting the sparse retrieval component benefited long synthetic questions but not short authentic ones, costing up to 8$\times$ the latency of the fastest configuration we tested. We propose treating synthetic and authentic query sets as complementary extremes of the query-quality spectrum: synthetic data verifies maximum retrieval capacity under idealized conditions, while authentic queries test system robustness to the imprecise, underspecified inputs of real users.
\end{abstract}

\ccsdesc[500]{Information systems~Retrieval models and ranking}
\ccsdesc[500]{Information systems~Question answering}
\ccsdesc[300]{Information systems~Evaluation of retrieval results}
\keywords{retrieval-augmented generation, RAG evaluation, synthetic benchmarks, query distribution shift, pre-deployment evaluation, hybrid retrieval}

\maketitle
\section{Introduction}
Retrieval-augmented generation (RAG) grounds language model responses in an external document corpus~\cite{lewis2020rag,gao2023rag_survey}. Before deployment, practitioners typically evaluate RAG pipelines on offline benchmarks comprising synthetic question--answer pairs generated from the target corpus. Synthetic queries are easy to generate from the target corpus and are used to optimize RAG systems, since they provide large, low-cost evaluation sets long before authentic user traffic becomes available. This convenience, however, makes it tempting to treat synthetic benchmarks as a proxy for real-world usage rather than as one of several complementary tests. Recent RAG evaluation frameworks no longer evaluate final answer accuracy in isolation; instead, they measure individual components such as the retriever and the language model separately~\cite{saadfalcon2024ares,ru2024ragchecker,filice2025datamorgana}.

While strong component-level benchmark performance is generally assumed to translate to strong post-deployment performance, this perspective overlooks the fact that synthetic and authentic queries represent two \emph{complementary extremes} of the query-quality spectrum. Synthetic generation is instructed to be diverse: the generator visits many documents and spreads questions broadly, phrasing them in full, well-structured sentences. The resulting benchmark verifies the system's fundamental capability to retrieve correct sources across the entire corpus for well-formed, document-grounded questions~\cite{filice2025datamorgana,zhu2025rageval}.

However, real users often arrive with specific, time-sensitive needs, producing a query distribution that is concentrated, repetitive, and focused on administrative and procedural tasks~\cite{page2017pounce,antico2024unimib,nguyenduc2025maraus}. Because gathering authentic, short queries at scale is notoriously difficult, we argue that evaluating a RAG system exclusively on synthetic data tests its theoretical capability, but fails to test its practical robustness.

We demonstrate this empirically with \emph{MiNIonek}, an administrative chatbot developed at the Faculty of Mathematics and Information Science, Warsaw University of Technology, comparing synthetic questions against authentic pre-deployment queries written by students. The two distributions diverge across intent distribution, question length, source coverage, and evaluability. The retrieval configuration that scores highest on the synthetic benchmark exhibits the largest performance gap on authentic queries. In our setting, \emph{BM25} (Best Matching 25) sparse retrieval~\cite{robertson2009bm25} benefits long, keyword-rich synthetic questions but contributes virtually nothing to short, keyword-like authentic queries.

Our contributions are:
\begin{itemize}
  \item[$\star$] We characterize the gap between synthetic and authentic distributions across five dimensions: intent distribution, question length, source coverage, evaluability, and retrieval metrics. The first four are observable before deployment (Section~\ref{sec:gap}).
  \item[$\star$] We demonstrate that synthetic RAG benchmarks alone can invert retrieval-system selection under realistic query distributions, leading engineering teams to make suboptimal architectural choices (Section~\ref{sec:results}).
  \item[$\star$] We provide an intentionally lightweight and reproducible pre-deployment checklist that treats synthetic and authentic queries as complementary tests, using their mismatch as a direct deployment-risk signal (Section~\ref{sec:discussion}).
\end{itemize}
\section{Related Work}
\paragraph{RAG evaluation.}
RAG evaluation has shifted from end-to-end accuracy toward component-level metrics. Frameworks like RAGAS~\cite{es2023ragas}, ARES~\cite{saadfalcon2024ares}, and RAGChecker~\cite{ru2024ragchecker} diagnose retrieval and generation separately using reference-free metrics such as faithfulness, answer relevance, context precision, and context recall. Recent benchmarks (MIRAGE~\cite{park2025mirage}, MEMERAG~\cite{cruzblandon2025memerag}, GaRAGe~\cite{sorodoc2025garage}) further broaden evaluation to multilinguality and long-form answers. While these improve measurement on existing test sets, our work is complementary: we show the test-set distribution itself represents a potential point of failure.
\paragraph{Synthetic test collection quality.}
Synthetic datasets reduce evaluation costs by generating scalable resources (ARES \cite{saadfalcon2024ares}, RAGEval \cite{zhu2025rageval}, YourBench \cite{shashidhar2025yourbench}). Tools like DataMorgana~\cite{filice2025datamorgana} even allow designers to configure expected question distributions for enterprise applications. However, these methods still rely on assumptions about expected future traffic. We argue that while a high-quality synthetic benchmark is a valuable coverage test, it should not be treated as a representative sample of real user traffic.
\paragraph{Retrieval strategy and query characteristics.}
Dense~\cite{karpukhin2020dpr} and sparse (\emph{BM25}~\cite{robertson2009bm25}) retrieval capture different signals. Hybrid retrieval combines both, but its benefit scales with query length and keyword density~\cite{chen2024benchmarking}. We demonstrate a practical consequence of this interaction: choosing hybrid over dense retrieval based solely on synthetic benchmarks can incur significant latency penalties with no practical benefit for short authentic queries.
\paragraph{Answerability and failure modes.}
Recent work recognizes that RAG systems must also be evaluated on their ability to reject or route unsupported questions. New benchmarks introduce taxonomies for unanswerable queries~\cite{peng2024uaeval4rag}, deflective responses~\cite{sorodoc2025garage}, and realistic failure modes like missing evidence or out-of-scope requests~\cite{liu2025crumq}. We highlight that synthetic benchmarks inherently omit these scenarios, leaving the system's refusal behaviors un\-tested.
\paragraph{Student-facing university chatbots.}
Higher-education chatbots have evolved from traditional FAQ systems~\cite{ranoliya2017faq,santana2021sandra} to RAG-based architectures. However, recent deployments highlight strict operational realities: omitted information in Unimib Assistant~\cite{antico2024unimib}, extensive human-handoff protocols in Pounce~\cite{page2017pounce}, rigid constraints against privacy leaks in 61A-Bot~\cite{zamfirescu2024bot}, and multi-agent orchestration for safe query processing in MARAUS~\cite{nguyenduc2025maraus}. These studies evaluate chatbot usability and architecture; we focus on the distributional mismatch between synthetic evaluation questions and authentic student needs.
\section{The Synthetic--Authentic Gap}
\label{sec:gap}
Synthetic queries approximate detailed, document-grounded usage; authentic survey queries approximate terse, low-effort, high-urgency usage. Table~\ref{tab:dimensions} summarizes five dimensions along which these distributions diverge. The first four can be measured before deployment: intent distribution and question length need no annotation at all, and source coverage and evaluability need only a coarse mapping from each query to a source page, not graded relevance judgments. The fifth, retrieval metrics, requires running a retriever.
\begin{table}[t]
\centering
\caption{Five dimensions of the synthetic--authentic gap, each with the risk it signals. The first four are measurable before deployment; the fifth requires running a retriever, and we report Hit@5 as its example. Values are from our case study (Section~\ref{sec:results}), over 1{,}851 synthetic and 322 authentic queries, except source coverage and Hit@5, which use the 165 authentic queries carrying a source URL. Gini is over the category shares, from 0 (uniform) to 1 (all questions in one category).}
\label{tab:dimensions}
\small
\setlength{\tabcolsep}{4pt}
\begin{tabularx}{\columnwidth}{@{}
  >{\raggedright\arraybackslash}p{1.6cm}
  >{\raggedright\arraybackslash}X
  >{\raggedright\arraybackslash}X
  >{\raggedright\arraybackslash}p{2.2cm}
  @{}}
\toprule
\textbf{Dimension} & \textbf{Synthetic} & \textbf{Authentic} & \textbf{Risk} \\
\midrule
Intent \newline distribution & Broad (Gini 0.37); \newline Top-3: 37.3\% & Concentrated (Gini 0.59); \newline Top-3: 59.0\% & Metrics overstate broad coverage \\ \addlinespace
Question \newline length & 15.7 words avg. & 6.8 words avg. & Harder retrieval; \emph{BM25} unhelpful \\ \addlinespace
Source \newline coverage & 165 unique sources \newline (Top-3: 15.1\%) & 53 unique sources \newline (Top-3: 44.4\%) & High-traffic docs dominate \\ \addlinespace
Evaluability & 100\% grounded & 49\% not directly \newline evaluable & Refusal behavior untested \\ \addlinespace
Retrieval \newline metrics (Hit@5) & 0.82--0.90 & 0.53--0.55 & Score not predictive \\
\bottomrule
\end{tabularx}
\end{table}

\textbf{Intent distribution.} Synthetic generators tend to spread questions across the entire corpus; the generator's defaults did so here. Real users cluster around immediate administrative tasks---checking class schedules, verifying exam deadlines, submitting forms. This gap reveals which parts of the corpus matter in practice compared to what the benchmark over-exercises.

{\emergencystretch=3em
\textbf{Question length.} Generating from sources inherently produces specific, well-formed questions. Real users write short, direct queries---often a single phrase. The qualitative difference is stark: a generated question might read \emph{``What administrative conditions must a student meet to register for the next semester?''} while students simply write \emph{``Session end date?''} or \emph{``Timetable?''}.
\par}

\textbf{Source coverage.} Because synthetic queries touch many documents, they act as a rigorous coverage test; authentic queries repeatedly hit a tight cluster of pages. A system can score highly on synthetic data while failing on the handful of documents users depend on most.

\textbf{Evaluability.} Synthetic questions are generated from existing documents and are answerable by definition. Authentic pre-deployment queries frequently concern volatile information not fully represented in a static corpus: nearly half require either live data sources or explicit refusal handling, neither of which standard synthetic benchmarks test.

\textbf{Retrieval metrics.} Retrieval metrics such as Hit@5---the share of queries whose relevant source appears in the top five results---are the standard measure of retrieval quality; Table~\ref{tab:dimensions} reports Hit@5 as the example. Part of a score can come from the query distribution alone: because authentic queries concentrate on a handful of pages, always returning the five most requested pages, whatever the question, already scores 0.545 on them, against 0.233 on the synthetic set. A retrieval metric alone therefore cannot say how much of a score is retrieval and how much is concentration, which is what the four pre-deployment dimensions are for.
\section{Case Study}
\label{sec:results}
\subsection{Study Design}
\emph{MiNIonek} answers student questions about timetables, scholarships, syllabi, and administrative procedures. The corpus consists of official faculty web pages and PDF documents---we call each indexed page or document a \emph{source}---embedded with \href{https://huggingface.co/BAAI/bge-m3}{\texttt{BAAI/bge-m3}}~\cite{xiao2024bge} and indexed in two widely-used open-source vector backends: \href{https://www.trychroma.com/}{ChromaDB} (for dense-only retrieval)
and \href{https://qdrant.tech/}{Qdrant} (which stores both dense vectors and sparse \emph{BM25} vectors for hybrid retrieval).

\textbf{Synthetic queries.} We used the default interface of Google's \href{https://notebook.google.com/}{Gemini Notebook}~\cite{notebooklm2024}, released and used here under its former name NotebookLM, as a representative practitioner workflow for rapid benchmark creation rather than a strictly controlled scientific generator. This yielded 1{,}851 questions drawn from 165 unique sources.

\textbf{Authentic queries.} Before deployment no usage logs exist, and even published deployments rarely release logs carrying ground-truth source mappings~\cite{page2017pounce,nguyenduc2025maraus}. We therefore collected questions via a structured survey asking enrolled students for queries they recently needed answered regarding their studies. After deduplication, we retained 322 authentic queries from 89 respondents.

\textbf{Annotation protocol.} Ground truth was established by two university domain experts: both study at the faculty, use its website routinely, and curated the document corpus during system construction, so they know what each indexed page contains and which administrative process every question refers to. Each question was labeled by one of the two annotators, with the set split between them. Because student queries name the administrative process they concern, the mapping to a source page is near-deterministic once the corpus is known; the annotators jointly reviewed the cases where a query touched more than one process. Queries were marked as ``not directly evaluable'' if they required personal context, live data not present in the corpus, or were too ambiguous to map to a single document. Only 165 queries had a valid static source URL for traditional retrieval evaluation; the remaining 157 queries (49\%) lacked a grounding document and were thus categorized as not directly evaluable.
\subsection{Distributional Comparison}

\begin{figure}[t]
\centering
\includegraphics[width=\columnwidth]{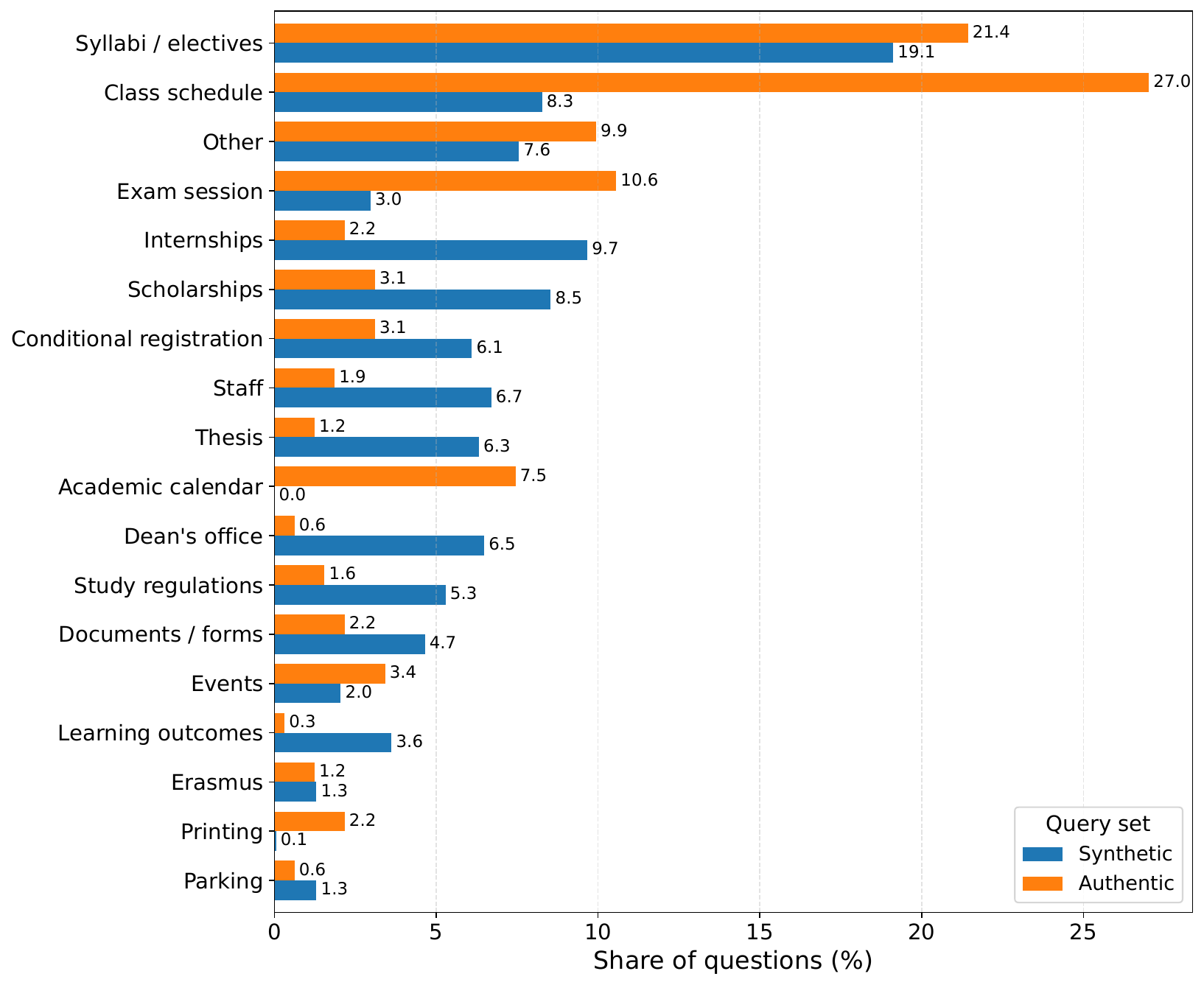}
\caption{Category distribution over all 1{,}851 synthetic (blue) and all 322 authentic (orange) queries, including the authentic ones with no static source, sorted by combined share.}
\label{fig:categories}
\end{figure}

\textbf{Category scheme.} The categories used in Table~\ref{tab:summary} and Figure~\ref{fig:categories} were induced from the data rather than taken from an existing taxonomy: the annotators open-coded the authentic queries by the student-facing task each expressed and reconciled their labels into a single scheme. The scheme was frozen before the synthetic set was labeled and then applied identically to both sets, so the categories stay anchored to real student needs rather than to the structure of the corpus.

Table~\ref{tab:summary} shows the structural disparities between the datasets. The synthetic set is evenly distributed across categories (Gini 0.37); the authentic set, conversely, is heavily concentrated around a few administrative and procedural intents (Gini 0.59). The Jensen--Shannon (JS) divergence~\cite{lin1991divergence} between the two category distributions is 0.203 bits, on a scale where 0 means identical distributions and 1 bit means no overlap at all. Question length diverges sharply: synthetic questions average 15.7 words compared to 6.8 for authentic queries. A Mann--Whitney $U$ test~\cite{mann1947test} confirms this length difference is statistically significant ($p < 0.001$) with a large effect size (Cliff's $\delta = 0.917$~\cite{cliff1993dominance}).

The most pronounced structural gap is source coverage. Synthetic queries draw from 165 sources, whereas authentic queries draw from only 53, and they concentrate heavily within those: a single source---the university academic calendar---accounts for 25.3\% of the evaluable authentic queries. The top three authentic sources carry 44.4\% of the traffic, compared to just 15.1\% for the top three synthetic sources.
\begin{table}[t]
\centering
\caption{Dataset-level comparison. Category and length statistics are over all 322 authentic queries; source statistics over the 165 that carry a source URL. A Mann--Whitney $U$ test on question length yields $p < 0.001$ with a large effect size (Cliff's $\delta = 0.917$).}
\label{tab:summary}
\small
\begin{tabularx}{\columnwidth}{@{} >{\raggedright\arraybackslash}X >{\raggedright\arraybackslash}p{2.3cm} >{\raggedright\arraybackslash}p{2.3cm} @{}}
\toprule
\textbf{Dimension} & \textbf{Synthetic} & \textbf{Authentic} \\
\midrule
Questions              & 1{,}851      & 322     \\
Unique categories      & 17           & 18      \\
Top category (share)   & Syllabi/electives (19.1\%) & Class schedule (27.0\%) \\
Avg.\ length (words)   & 15.7         & 6.8     \\
Median length (words)  & 15           & 6       \\
Unique sources    & 165          & 53      \\
Top source share       & 5.5\%        & 25.3\%  \\
Top-3 sources share    & 15.1\%       & 44.4\%  \\
Not directly evaluable & 0\%          & 49\%    \\
\bottomrule
\end{tabularx}
\end{table}

Figure~\ref{fig:categories} breaks down the categories. \emph{Class schedule} dominates authentic queries at 27.0\% but ranks fourth in the synthetic set at 8.3\%. \emph{Academic calendar} (7.5\% authentic) has no synthetic counterpart. Conversely, generation over-represents \emph{Internships} by a factor of four (9.7\% vs.\ 2.2\% authentic) and \emph{Thesis} procedures by five (6.3\% vs.\ 1.2\%).

\begin{figure}[t]
\centering
\includegraphics[width=\columnwidth]{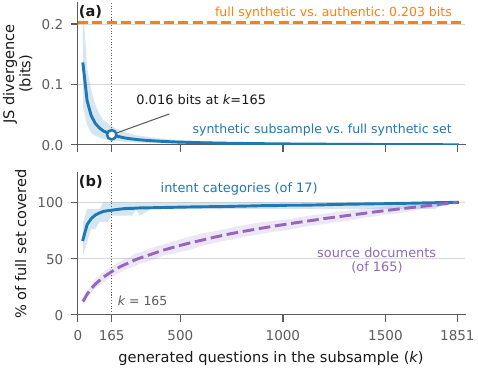}
\caption{\textbf{Does the size of the synthetic set explain the gap?} At each $k$ we draw 1{,}000 random subsamples of $k$ questions from the 1{,}851 synthetic ones; shaded bands are 95\% ranges over those draws. (a) JS divergence between a subsample's category distribution and that of the \emph{full} synthetic set; the horizontal line is a constant, not a curve, and does not depend on $k$. (b) How much of the full synthetic set a subsample covers on two dimensions from Table~\ref{tab:dimensions}: its 17 intent categories are almost entirely covered within a few hundred questions, whereas source coverage grows steadily and is complete only at $k = 1{,}851$.}
\label{fig:saturation}
\end{figure}

\textbf{Effect of set size.} The two sets differ in size---1{,}851 synthetic against the 165 evaluable authentic queries---because they differ in cost: synthetic questions are essentially free, authentic ones bounded by how many students respond. Equalizing them changes neither the category-distribution gap nor the retrieval gap. Figure~\ref{fig:saturation}(a) draws 1{,}000 random subsamples of $k$ synthetic questions at each size $k$ and measures the JS divergence between a subsample's category distribution and that of the full synthetic set. At the size-matched point $k = 165$ it is 0.016 bits (95\% range 0.007--0.031), an order of magnitude below the 0.203 bits separating the full synthetic set from the authentic one. The same holds for retrieval: re-scoring 1{,}000 size-165 subsamples gives Hit@5 of $0.896 \pm 0.023$ for the hybrid configuration, never below 0.824, against 0.527 on the evaluable authentic queries. Figure~\ref{fig:saturation}(b) shows why the synthetic set is so insensitive to its own size. Intent coverage is nearly complete early---165 questions already reach 93\% of the 17 categories the full set ever visits---while source coverage keeps climbing with $k$. Extra generation therefore buys breadth across the corpus rather than new student intents: it spreads questions over more documents, while authentic demand stays concentrated on 53 pages.

\subsection{Retrieval Performance}
Table~\ref{tab:retrieval} details standard information retrieval metrics. The $\Delta$Hit@5 column indicates the performance drop moving from synthetic to authentic queries. To capture uncertainty, we report 95\% confidence intervals for $\Delta$Hit@5 computed via 10,000 bootstrap resamples. Following recent diagnostic RAG evaluation work, we report retrieval metrics separately from answer-generation metrics because retrieval and generation failures can arise independently~\cite{ru2024ragchecker,saadfalcon2024ares,nguyenduc2025maraus}.
\begin{table*}[t]
\centering
\caption{Retrieval results on the synthetic (1{,}851) and authentic (165 questions) query sets; all configurations use \texttt{BAAI/bge-m3}. $\Delta$Hit@5 is Hit@5$_{\text{authentic}}$ $-$ Hit@5$_{\text{synthetic}}$, with 95\% CIs from 10,000 bootstrap resamples; a larger negative value means the configuration is more misleading about deployment readiness. Best synthetic score per column and lowest latency are \underline{underlined}; the largest drop is in \textbf{bold}.}
\label{tab:retrieval}
\setlength{\tabcolsep}{4.5pt}
\begin{tabular}{llccccccc}
\toprule
\textbf{Configuration} & \textbf{Query set}
  & \textbf{Hit@1} & \textbf{Hit@3} & \textbf{Hit@5} & \textbf{Hit@10}
  & \textbf{MRR@10}
  & \textbf{$\Delta$Hit@5 [95\% CI]}
  & \textbf{Latency / query (s)} \\
\midrule
ChromaDB (dense)
  & Synthetic & 0.532 & 0.786 & 0.820 & 0.843 & 0.662 & \multirow{2}{*}{$-$0.293~[$-$0.37, $-$0.22]} & \multirow{2}{*}{\underline{0.33}} \\
  & Authentic & 0.321 & 0.461 & 0.527 & 0.576 & 0.400 & & \\
\midrule
Qdrant (dense)
  & Synthetic & 0.555 & 0.814 & 0.849 & 0.873 & 0.687 & \multirow{2}{*}{$-$0.304~[$-$0.38, $-$0.23]} & \multirow{2}{*}{1.57} \\
  & Authentic & 0.315 & 0.497 & 0.545 & 0.582 & 0.413 & & \\
\midrule
Qdrant (hybrid)
  & Synthetic & \underline{0.566} & \underline{0.858} & \underline{0.896} & \underline{0.910} & \underline{0.709} & \multirow{2}{*}{$-$\textbf{0.369}~[$-$0.44, $-$0.29]} & \multirow{2}{*}{2.70} \\
  & Authentic & 0.321 & 0.461 & 0.527 & 0.576 & 0.400 & & \\
\bottomrule
\end{tabular}
\end{table*}

The results show three clear patterns. First, the gap is large, dropping 29--37 percentage points in Hit@5 across all configurations. A Cochran--Mantel--Haenszel test~\cite{mantel1959statistical} confirms that this performance drop remains significant after controlling for category ($p < 0.001$), with $\Delta$Hit@5 negative in 6 of the 7 largest categories present in both sets. The sole exception is \emph{Class schedule}, where authentic queries use broad keywords that easily match the main schedule page, whereas synthetic generation creates overly complex queries that the retriever struggles to resolve. Second, Qdrant hybrid is the highest-performing configuration on the synthetic benchmark but simultaneously experiences the largest drop ($\Delta$Hit@5 = $-$0.369). Third, performance across the three configurations is virtually identical on authentic queries. Because \emph{BM25} matches on exact keywords, it favors the long synthetic queries and adds nothing to short authentic ones: within the same backend, hybrid retrieval scores 0.527 on authentic queries against 0.545 for dense alone, at 1.7$\times$ the latency (2.70\,s versus 1.57\,s per query), and 8$\times$ that of the fastest configuration. Choosing it on synthetic evidence therefore buys latency and no accuracy, which matters where response times are constrained~\cite{zamfirescu2024bot,xu2025ragops}.
\subsection{Evaluability}
49\% of authentic queries (157 of 322) could not be mapped to a single static source URL. Evaluability is not uniform across intents: the two most-asked categories, \emph{Class schedule} and \emph{Syllabi/electives}, can be mapped to a single static source in only 47\% and 30\% of cases, whereas every \emph{Academic calendar} query can: the benchmark is strongest where demand is weakest. This aligns with recent work arguing that systems should be tested on whether they appropriately reject, defer, or route unsupported requests rather than assuming every query is answerable from the indexed corpus~\cite{peng2024uaeval4rag}. Missing this capability leads to errors in deployment: Unimib Assistant generated unclickable links and omitted crucial information when forced to answer outside its knowledge~\cite{antico2024unimib}, and Pounce showed that high-stakes unanswerable queries require clear human handoff~\cite{page2017pounce}.

\section{Discussion and Limitations}
\label{sec:discussion}
\paragraph{Implications.}
Our results are not an argument against synthetic evaluation. Generated query sets stress-test whether a RAG system can retrieve broadly across the indexed corpus under well-specified, source-grounded information needs, and in that role they are an efficient coverage test and a useful sanity check before deployment.

The problem arises when such benchmarks are read as representative samples of future traffic. Authentic pre-deployment queries test a different capability: robustness to short, underspecified, repetitive, operational requests, including cases where the correct behavior is refusal, clarification, or human handoff rather than retrieval. Deployment readiness requires inspecting both extremes, and treating the mismatch between them as a risk signal rather than as a reason to discard either benchmark.
\paragraph{Pre-deployment checklist.}
Based on our results, we propose the following steps before deploying any RAG architecture:
\begin{enumerate}
  \item Generate a synthetic question set to validate broad retrieval capability and pipeline coverage~\cite{filice2025datamorgana}.
  \item Collect 50--100 authentic queries from intended users or domain proxies (no formal annotation required) to test robustness.
  \item Compare the synthetic and authentic sets across intent distribution, question length, source coverage, and evaluability.
  \item Finalize the retrieval architecture based on the authentic query profile, balancing latency trade-offs rather than optimizing solely for the synthetic benchmark~\cite{xu2025ragops}.
  \item Implement strict refusal policies and human-handoff triggers for unanswerable queries~\cite{zamfirescu2024bot}.
\end{enumerate}
\paragraph{Limitations.}
While authentic survey queries are closer to low-effort, pre-deployment user behavior, they are not necessarily a complete proxy for live traffic. The extreme brevity of our authentic queries may also be exacerbated by survey fatigue, since respondents invest less effort in a survey than in conversation with an agent. The 322 queries come from 89 respondents and are therefore not fully independent, whereas the tests we report treat them as such. Our synthetic generation also relies on the default Gemini Notebook interface, representing one practitioner workflow rather than a parameterized scientific generator.
\section{Conclusion}
Synthetic benchmarks measure retrieval coverage under idealized, source-grounded conditions; authentic queries measure robustness under underspecified, operational conditions. Neither alone is sufficient for deployment-readiness assessment. Generated sets reliably confirm whether a RAG system can retrieve broadly across its corpus, but they obscure the reality that authentic queries are concentrated, short, and frequently demand context beyond a static document collection. Integrating even a small set of authentic queries exposes risks that synthetic evaluation alone leaves unexamined. The ranking itself inverts: the configuration that leads on every synthetic metric (Hit@5 0.896) is not the one that leads on authentic queries (0.527, against 0.545 for dense retrieval alone), so a synthetic-only comparison selects the slower architecture.
\begin{acks}
We sincerely thank Dr.\ Anna Wr\'oblewska of Warsaw University of Technology for her invaluable guidance, insightful feedback, and continuous support throughout this work.
F.J.K. was funded by the European Union under the Horizon Europe project OMINO (grant agreement No.~101086321). Views and opinions expressed are those of the authors alone and do not necessarily reflect those of the European Union or the European Research Executive Agency. Neither the European Union nor the European Research Executive Agency can be held responsible for them. F.J.K. was also co-financed with funds from the Polish Ministry of Science and Higher Education under the programme entitled ``International Co-Financed Projects''.
\end{acks}
\section*{GenAI Usage Disclosure}
Following ACM's guidelines on the use of generative AI tools, we disclose that such tools were used solely for grammar checking (Grammarly) and for assistance with code debugging during the preparation of this paper. All research ideas, experiments, analysis, and writing were conducted and critically reviewed by the authors. No part of the scientific content or creative reasoning has been generated or substantially rewritten by generative AI tools.
\bibliographystyle{ACM-Reference-Format}
\balance
\bibliography{references}

@inproceedings{lewis2020rag,
author = {Lewis, Patrick and Perez, Ethan and Piktus, Aleksandra and Petroni, Fabio and Karpukhin, Vladimir and Goyal, Naman and K{\"u}ttler, Heinrich and Lewis, Mike and Yih, Wen-tau and Rockt{\"a}schel, Tim and Riedel, Sebastian and Kiela, Douwe},
title = {Retrieval-augmented generation for knowledge-intensive NLP tasks},
pages     = {9459--9474},
year = {2020},
isbn = {9781713829546},
publisher = {Curran Associates Inc.},
address = {Red Hook, NY, USA},
booktitle = {Proceedings of the 34th International Conference on Neural Information Processing Systems},
articleno = {793},
numpages = {16},
location = {Vancouver, BC, Canada},
series = {NIPS '20}
}

@article{gao2023rag_survey,
  title   = {Retrieval-Augmented Generation for Large Language Models: A Survey},
  author  = {Gao, Yunfan and Xiong, Yun and Gao, Xinyu and Jia, Kangxiang and Pan, Jinliu and Bi, Yuxi and Dai, Yi and Sun, Jiawei and Wang, Haofen},
  journal = {arXiv preprint arXiv:2312.10997},
  year={2024},
  eprint={2312.10997},
  archivePrefix={arXiv},
  primaryClass={cs.CL},
  url={https://arxiv.org/abs/2312.10997}, 
}

@inproceedings{es2023ragas,
    title = "{RAGA}s: Automated Evaluation of Retrieval Augmented Generation",
    author = "Es, Shahul  and
      James, Jithin  and
      Espinosa Anke, Luis  and
      Schockaert, Steven",
    editor = "Aletras, Nikolaos  and
      De Clercq, Orphee",
    booktitle = "Proceedings of the 18th Conference of the European Chapter of the Association for Computational Linguistics: System Demonstrations",
    month = mar,
    year = "2024",
    address = "St. Julians, Malta",
    publisher = "Association for Computational Linguistics",
    url = "https://aclanthology.org/2024.eacl-demo.16/",
    doi = "10.18653/v1/2024.eacl-demo.16",
    pages = "150--158",
}

@inproceedings{karpukhin2020dpr,
    title = "Dense Passage Retrieval for Open-Domain Question Answering",
    author = "Karpukhin, Vladimir  and
      Oguz, Barlas  and
      Min, Sewon  and
      Lewis, Patrick  and
      Wu, Ledell  and
      Edunov, Sergey  and
      Chen, Danqi  and
      Yih, Wen-tau",
    editor = "Webber, Bonnie  and
      Cohn, Trevor  and
      He, Yulan  and
      Liu, Yang",
    booktitle = "Proceedings of the 2020 Conference on Empirical Methods in Natural Language Processing (EMNLP)",
    month = nov,
    year = "2020",
    address = "Online",
    publisher = "Association for Computational Linguistics",
    url = "https://aclanthology.org/2020.emnlp-main.550/",
    doi = "10.18653/v1/2020.emnlp-main.550",
    pages = "6769--6781"
}

@article{robertson2009bm25,
  title   = {The Probabilistic Relevance Framework: {BM25} and Beyond},
  author  = {Robertson, Stephen and Zaragoza, Hugo},
  journal = {Foundations and Trends in Information Retrieval},
  volume  = {3},
  number  = {4},
  pages   = {333--389},
  year    = {2009}
}

@inproceedings{xiao2024bge,
    title = "{M}3-Embedding: Multi-Linguality, Multi-Functionality, Multi-Granularity Text Embeddings Through Self-Knowledge Distillation",
    author = "Chen, Jianlyu  and
      Xiao, Shitao  and
      Zhang, Peitian  and
      Luo, Kun  and
      Lian, Defu  and
      Liu, Zheng",
    editor = "Ku, Lun-Wei  and
      Martins, Andre  and
      Srikumar, Vivek",
    booktitle = "Findings of the Association for Computational Linguistics: ACL 2024",
    month = aug,
    year = "2024",
    address = "Bangkok, Thailand",
    publisher = "Association for Computational Linguistics",
    url = "https://aclanthology.org/2024.findings-acl.137/",
    doi = "10.18653/v1/2024.findings-acl.137",
    pages = "2318--2335"
}

@inproceedings{chen2024benchmarking,
author = {Chen, Jiawei and Lin, Hongyu and Han, Xianpei and Sun, Le},
title = {Benchmarking large language models in retrieval-augmented generation},
year = {2024},
isbn = {978-1-57735-887-9},
publisher = {AAAI Press},
url = {https://doi.org/10.1609/aaai.v38i16.29728},
doi = {10.1609/aaai.v38i16.29728},
booktitle = {Proceedings of the Thirty-Eighth AAAI Conference on Artificial Intelligence and Thirty-Sixth Conference on Innovative Applications of Artificial Intelligence and Fourteenth Symposium on Educational Advances in Artificial Intelligence},
articleno = {1980},
numpages = {9},
series = {AAAI'24/IAAI'24/EAAI'24}
}

@misc{notebooklm2024,
  title        = {{Gemini Notebook}}, author       = {{Google}},
  howpublished = {\url{https://notebook.google.com/}}, year         = {2026},
  note         = {Formerly NotebookLM; renamed 16 July 2026}
}

@misc{antico2024unimib,
  title={Unimib Assistant: designing a student-friendly RAG-based chatbot for all their needs},
  author={Antico, Chiara and Giordano, Stefano and Koyuturk, Cansu and Ognibene, Dimitri},
  year={2024},
  eprint={2411.19554},
  archivePrefix={arXiv},
  primaryClass={cs.HC},
  doi={10.48550/arXiv.2411.19554}
}

@inproceedings{zamfirescu2024bot,
author = {Zamfirescu-Pereira, J.D. and Qi, Laryn and Hartmann, Bj{\"o}rn and DeNero, John and Norouzi, Narges},
title = {61A Bot Report: AI Assistants in CS1 Save Students Homework Time and Reduce Demands on Staff. (Now What?)},
year = {2025},
isbn = {9798400705311},
publisher = {Association for Computing Machinery},
address = {New York, NY, USA},
url = {https://doi.org/10.1145/3641554.3701864},
doi = {10.1145/3641554.3701864},
booktitle = {Proceedings of the 56th ACM Technical Symposium on Computer Science Education V. 1},
pages = {1309–1315},
numpages = {7},
location = {Pittsburgh, PA, USA},
series = {SIGCSETS 2025}
}

@misc{nguyenduc2025maraus,
  title={An Empirical Study of Multi-Agent RAG for Real-World University Admissions Counseling},
  author={Nguyen-Duc, Anh and Manh, Chien Vu and Tran, Bao Anh and Ngo, Viet Phuong and Le Chi, Luan and Nguyen, Anh Quang},
  year={2025},
  eprint={2507.11272},
  archivePrefix={arXiv},
  primaryClass={cs.SE},
  doi={10.48550/arXiv.2507.11272}
}

@misc{peng2024uaeval4rag,
  title={Unanswerability Evaluation for Retrieval Augmented Generation},
  author={Peng, Xiangyu and Choubey, Prafulla Kumar and Xiong, Caiming and Wu, Chien-Sheng},
  year={2024},
  eprint={2412.12300},
  archivePrefix={arXiv},
  doi={10.48550/arXiv.2412.12300}
}

@misc{xu2025ragops,
  title={RAGOps: Operating and Managing Retrieval-Augmented Generation Pipelines},
  author={Xu, Xiwei and Weytjens, Hans and Zhang, Dawen and Lu, Qinghua and Weber, Ingo and Zhu, Liming},
  year={2025},
  eprint={2506.03401},
  archivePrefix={arXiv},
  doi={10.48550/arXiv.2506.03401}
}

@article{page2017pounce,
title = {How an Artificially Intelligent Virtual Assistant Helps Students Navigate the Road to College},
author = {Lindsay C. Page and Hunter Gehlbach},
journal = {AERA Open}, volume = {3}, number = {4}, pages = {2332858417749220},
year = {2017}, doi = {10.1177/2332858417749220},
URL = {https://doi.org/10.1177/2332858417749220}
}

@inproceedings{santana2021sandra,
  title = {A Chatbot to Support Basic Students Questions},
  author = {Santana, Rafael and Ferreira, Saulo and Rolim, Vitor and Miranda, P{\'e}ricles and Nascimento, Andr{\'e} and Mello, Rafael Ferreira},
  booktitle = {Proceedings of the IV Latin American Conference on Learning Analytics (LALA 2021)},
  series = {CEUR Workshop Proceedings}, volume = {3059}, pages = {58--67},
  year = {2021}, address = {Arequipa, Peru}, publisher = {CEUR-WS.org}
}

@inproceedings{ranoliya2017faq,
author={Ranoliya, Bhavika R. and Raghuwanshi, Nidhi and Singh, Sanjay},
  booktitle={2017 International Conference on Advances in Computing, Communications and Informatics (ICACCI)}, 
  title={Chatbot for university related FAQs}, 
  year={2017},
  volume={},
  number={},
  pages={1525-1530},
  doi={10.1109/ICACCI.2017.8126057},
  publisher = {IEEE},
  address = {Udupi, India}
}

@inproceedings{saadfalcon2024ares,
    title = "{ARES}: An Automated Evaluation Framework for Retrieval-Augmented Generation Systems",
    author = "Saad-Falcon, Jon  and
      Khattab, Omar  and
      Potts, Christopher  and
      Zaharia, Matei",
    editor = "Duh, Kevin  and
      Gomez, Helena  and
      Bethard, Steven",
    booktitle = "Proceedings of the 2024 Conference of the North American Chapter of the Association for Computational Linguistics: Human Language Technologies (Volume 1: Long Papers)",
    month = jun,
    year = "2024",
    address = "Mexico City, Mexico",
    publisher = "Association for Computational Linguistics",
    url = "https://aclanthology.org/2024.naacl-long.20/",
    doi = "10.18653/v1/2024.naacl-long.20",
    pages = "338--354"
}

@inproceedings{ru2024ragchecker,
author = {Ru, Dongyu and Qiu, Lin and Hu, Xiangkun and Zhang, Tianhang and Shi, Peng and Chang, Shuaichen and Jiayang, Cheng and Wang, Cunxiang and Sun, Shichao and Li, Huanyu and Zhang, Zizhao and Wang, Binjie and Jiang, Jiarong and He, Tong and Wang, Zhiguo and Liu, Pengfei and Zhang, Yue and Zhang, Zheng},
title = {RAGCHECKER: a fine-grained framework for diagnosing retrieval-augmented generation},
year = {2024}, isbn = {9798331314385},
publisher = {Curran Associates Inc.}, address = {Red Hook, NY, USA},
booktitle = {Proceedings of the 38th International Conference on Neural Information Processing Systems},
articleno = {692}, numpages = {29}, location = {Vancouver, BC, Canada}, series = {NIPS '24}
}

@inproceedings{zhu2025rageval,
    title = "{RAGE}val: Scenario Specific {RAG} Evaluation Dataset Generation Framework",
    author = "Zhu, Kunlun  and
      Luo, Yifan  and
      Xu, Dingling  and
      Yan, Yukun  and
      Liu, Zhenghao  and
      Yu, Shi  and
      Wang, Ruobing  and
      Wang, Shuo  and
      Li, Yishan  and
      Zhang, Nan  and
      Han, Xu  and
      Liu, Zhiyuan  and
      Sun, Maosong",
    editor = "Che, Wanxiang  and
      Nabende, Joyce  and
      Shutova, Ekaterina  and
      Pilehvar, Mohammad Taher",
    booktitle = "Proceedings of the 63rd Annual Meeting of the Association for Computational Linguistics (Volume 1: Long Papers)",
    month = jul,
    year = "2025",
    address = "Vienna, Austria",
    publisher = "Association for Computational Linguistics",
    url = "https://aclanthology.org/2025.acl-long.418/",
    doi = "10.18653/v1/2025.acl-long.418",
    pages = "8520--8544",
    ISBN = "979-8-89176-251-0"
}

@inproceedings{filice2025datamorgana,
    title = "Generating {Q}{\&}{A} Benchmarks for {RAG} Evaluation in Enterprise Settings",
    author = "Filice, Simone  and
      Horowitz, Guy  and
      Carmel, David  and
      Karnin, Zohar  and
      Lewin-Eytan, Liane  and
      Maarek, Yoelle",
    editor = "Rehm, Georg  and
      Li, Yunyao",
    booktitle = "Proceedings of the 63rd Annual Meeting of the Association for Computational Linguistics (Volume 6: Industry Track)",
    month = jul,
    year = "2025",
    address = "Vienna, Austria",
    publisher = "Association for Computational Linguistics",
    url = "https://aclanthology.org/2025.acl-industry.33/",
    doi = "10.18653/v1/2025.acl-industry.33",
    pages = "469--484",
    ISBN = "979-8-89176-288-6"
}

@inproceedings{park2025mirage,
    title = "{MIRAGE}: A Metric-Intensive Benchmark for Retrieval-Augmented Generation Evaluation",
    author = "Park, Chanhee  and
      Moon, Hyeonseok  and
      Park, Chanjun  and
      Lim, Heuiseok",
    editor = "Chiruzzo, Luis  and
      Ritter, Alan  and
      Wang, Lu",
    booktitle = "Findings of the Association for Computational Linguistics: NAACL 2025",
    month = apr,
    year = "2025",
    address = "Albuquerque, New Mexico",
    publisher = "Association for Computational Linguistics",
    url = "https://aclanthology.org/2025.findings-naacl.157/",
    doi = "10.18653/v1/2025.findings-naacl.157",
    pages = "2883--2900",
    ISBN = "979-8-89176-195-7"
}

@inproceedings{sorodoc2025garage,
    title = "{G}a{RAG}e: A Benchmark with Grounding Annotations for {RAG} Evaluation",
    author = {Sorodoc, Ionut Teodor and Ribeiro, Leonardo F. R. and Blloshmi, Rexhina and Davis, Christopher and de Gispert, Adri{\`a}},
    editor = "Che, Wanxiang  and Nabende, Joyce  and Shutova, Ekaterina  and Pilehvar, Mohammad Taher",
    booktitle = "Findings of the Association for Computational Linguistics: ACL 2025",
    month = jul, year = "2025", address = "Vienna, Austria",
    publisher = "Association for Computational Linguistics",
    url = "https://aclanthology.org/2025.findings-acl.875/",
    doi = "10.18653/v1/2025.findings-acl.875", pages = "17030--17049", ISBN = "979-8-89176-256-5"
}

@inproceedings{cruzblandon2025memerag,
    title = "{MEMERAG}: A Multilingual End-to-End Meta-Evaluation Benchmark for Retrieval Augmented Generation",
    author = "Cruz Bland{\'o}n, Mar{\'i}a Andrea  and
      Talur, Jayasimha  and
      Charron, Bruno  and
      Liu, Dong  and
      Mansour, Saab  and
      Federico, Marcello",
    editor = "Che, Wanxiang  and
      Nabende, Joyce  and
      Shutova, Ekaterina  and
      Pilehvar, Mohammad Taher",
    booktitle = "Proceedings of the 63rd Annual Meeting of the Association for Computational Linguistics (Volume 1: Long Papers)",
    month = jul,
    year = "2025",
    address = "Vienna, Austria",
    publisher = "Association for Computational Linguistics",
    url = "https://aclanthology.org/2025.acl-long.1101/",
    doi = "10.18653/v1/2025.acl-long.1101",
    pages = "22577--22595",
    ISBN = "979-8-89176-251-0"
}

@InProceedings{liu2025crumq,
author="Liu, Gabrielle Kaili-May
and Li, Bryan
and Cohan, Arman
and Walden, William Gantt
and Yang, Eugene",
editor="Campos, Ricardo
and Jatowt, Adam
and Lan, Yanyan
and Aliannejadi, Mohammad
and Bauer, Christine
and MacAvaney, Sean
and Anand, Avishek
and Ren, Zhaochun
and Verberne, Suzan
and Bai, Nan
and Mansoury, Masoud",
title="Investigating Retrieval-Augmented Generation Systems on Unanswerable, Uncheatable, Realistic, Multi-hop Queries",
booktitle="Advances in Information Retrieval",
year="2026",
publisher="Springer Nature Switzerland",
address="Cham",
pages="360--370",
isbn="978-3-032-21300-6"
}

@article{cliff1993dominance,
  title = {Dominance Statistics: Ordinal Analyses to Answer Ordinal Questions},
  author = {Cliff, Norman},
  journal = {Psychological Bulletin},
  volume = {114},
  number = {3},
  pages = {494--509},
  year = {1993},
  publisher = {American Psychological Association},
  doi = {10.1037/0033-2909.114.3.494}
}

@ARTICLE{lin1991divergence,
  author={Lin, J.},
  journal={IEEE Transactions on Information Theory}, 
  title={Divergence measures based on the Shannon entropy}, 
  year={1991},
  volume={37},
  number={1},
  pages={145-151},
  doi={10.1109/18.61115}}

@article{mann1947test,
  title = {On a Test of Whether One of Two Random Variables is
           Stochastically Larger than the Other},
  author = {Mann, Henry B. and Whitney, Donald R.},
  journal = {The Annals of Mathematical Statistics},
  volume = {18},
  number = {1},
  pages = {50--60},
  year = {1947},
  publisher = {Institute of Mathematical Statistics},
  doi = {10.1214/aoms/1177730491}
}

@article{mantel1959statistical,
    author = {Mantel, Nathan and Haenszel, William},
    title = {Statistical Aspects of the Analysis of Data From Retrospective Studies of Disease},
    journal = {JNCI: Journal of the National Cancer Institute},
    volume = {22}, number = {4}, pages = {719-748}, year = {1959}, month = {04},
    doi = {10.1093/jnci/22.4.719}, url = {https://doi.org/10.1093/jnci/22.4.719},
}

@misc{shashidhar2025yourbench,
      title={YourBench: Easy Custom Evaluation Sets for Everyone}, 
      author={Sumuk Shashidhar and Clémentine Fourrier and Alina Lozovskia and Thomas Wolf and Gokhan Tur and Dilek Hakkani-Tür},
      year={2025},
      eprint={2504.01833},
      archivePrefix={arXiv},
      primaryClass={cs.CL},
      url={https://arxiv.org/abs/2504.01833}, 
}
\end{document}